\documentclass[10pt]{article}
\usepackage{fullpage}

\usepackage{geometry}
\usepackage{cite}
\usepackage{amsmath,graphicx}
\usepackage{amsmath,amssymb,amsfonts,bm,ntheorem}
\usepackage{graphicx}
\usepackage{textcomp}
\usepackage{xcolor}
\usepackage[small,bf]{caption}
\usepackage{color}
\usepackage{multicol}
\usepackage{rotating}
\usepackage{subfigure}
\usepackage{tablefootnote}
\usepackage[flushleft]{threeparttable}
\usepackage{rotating}
\usepackage{listings}
\usepackage{mathtools}

\usepackage{algorithm}
\usepackage{algpseudocode}

\algtext*{EndFor}

\usepackage{booktabs}
\usepackage{multirow}       
\usepackage{tabularx}       
\usepackage{hhline}
\usepackage{arydshln}		
\usepackage{xcolor}

\newcolumntype{L}[1]{>{\raggedright\arraybackslash}p{#1}}
\newcolumntype{C}[1]{>{\centering\arraybackslash}p{#1}}
\newcolumntype{R}[1]{>{\raggedleft\arraybackslash}p{#1}}
\newcolumntype{M}[1]{>{\centering\arraybackslash}m{#1}}

\renewcommand{\omega}{\phi}
\renewcommand{\Omega}{\Phi}

\def\prox{\mathsf{prox}}

\newcommand{\argmin}{\mathop{\rm argmin}}

\renewcommand{\leq}{\leqslant}
\renewcommand{\geq}{\geqslant}

\def\Dsf{{\mathsf{D}}}

\def\xbm{{\bm{x}}}

\def\ybm{{\bm{y}}}
\def\zbm{{\bm{z}}}

\def\Abm{{\bm{A}}}

\def\Tsf{\mathsf{T}}

\def\xbmhat{{\widehat{\bm{x}}}}

\def\R{\mathbb{R}}

\def\argmin{\mathop{\mathsf{arg\,min}}}

\def\prox{\mathsf{prox}}

\graphicspath{ {./images/} }

\newtheorem{theorem}{Theorem}

\begin{document}
\title{Bregman Plug-and-Play Priors} 
{\normalsize\author{Abdullah~H.~Al-Shabili\(^{1,}\)\thanks{Abdullah~H.~Al-Shabil and Xiaojian~Xu contributed equally to this work.  This work was supported in part by the NSF Award CCF-2043134.} , Xiaojian~Xu\(^{2, \ast}\), Ivan~Selesnick\(^1\), and Ulugbek~S.~Kamilov\(^{2, 3}\) \\
		\emph{\small \(^1 \) Department of Electrical and Computer Engineering, Tandon School of Engineering, New York University, USA.}\\
		\emph{\small \(^2 \) Department of Computer Science and Engineering,~Washington University in St.~Louis, MO 63130, USA.}\\
		\emph{\small \(^3 \) Department of Electrical and Systems Engineering,~Washington University in St.~Louis, MO 63130, USA.}\\
}}
\maketitle

\begin{abstract}
The past few years have seen a surge of activity around integration of deep learning networks and optimization algorithms for solving inverse problems. Recent work on plug-and-play priors (PnP), regularization by denoising (RED), and deep unfolding has shown the state-of-the-art performance of such integration in a variety of applications. However, the current paradigm for designing such algorithms is inherently Euclidean, due to the usage of the quadratic norm within the projection and proximal operators. We propose to broaden this perspective by considering a non-Euclidean setting based on the more general Bregman distance. Our new Bregman Proximal Gradient Method variant of PnP (PnP-BPGM) and Bregman Steepest Descent variant of RED (RED-BSD) replace the traditional updates in PnP and RED from the quadratic norms to more general Bregman distance. We present a theoretical convergence result for PnP-BPGM and demonstrate the effectiveness of our algorithms on Poisson linear inverse problems.
\end{abstract}


\section{Introduction} \label{sec:Introduction}
The recovery of an unknown signal $\xbm \in \R^n$ from its noisy measurements $\ybm \in \R^m$ can often be formulated as an inverse problem 
\begin{align}
\label{obse_model}
\ybm = \mathcal{N}(\Abm \, \xbm), 
\end{align}
\noindent
where $\Abm \in \mathbb{R}^{m \times n}$ is a measurement operator and $\mathcal{N}$ models the corruption of the measurements by noise, which could be signal dependent (e.g., Poisson noise) or signal independent (e.g., Gaussian noise). The solution of ill-posed inverse problems is often formulated as an optimization problem
\begin{align}
\label{invers_prob}
\xbmhat  = \argmin_{\xbm \in \mathbb{R}^n} \, f(\xbm) + g(\xbm)
\end{align}
\noindent where $f$ is the data-fidelity term and $g$ is the regularizer.

The past few years have seen a surge of efforts to integrate deep learning (DL) priors into iterative algorithms \cite{ongie2020deep, monga2021algorithm}. \emph{Plug-and-play priors (PnP)} \cite{venkatakrishnan2013plug} and \emph{regularization by denoising (RED)} \cite{romano2017little} are two methods that integrate pre-trained DL denoisers into iterative algorithms. Deep unfolding is a related strategy based on unfolding an iterative algorithm and including trainable blocks within it \cite{gregor2010learning}. Compared to the black-box DL, model-based DL methods integrate the physics-based knowledge of the measurement model. Their empirical success \cite{sreehari2016plug, liu2020rare}, has spurred a number of algorithmic extensions \cite{kamilov2017plug, ono2017primal, al2020learning}, as well as theoretical convergence analyses \cite{chen2018theoretical, ryu2019plug, sun2019online}. 

Most of the current work in PnP is fundamentally based on the traditional definition of the proximal operator that relies on the squared Euclidean norm. Under this definition the proximal operator can be naturally interpreted as the Gaussian denoiser. 
In this paper, we seek to broaden the family of PnP algorithms to the non-Euclidean setting by building on the recent work on Bregman proximal algorithms \cite{bauschke2017descent, lu2018relatively, teboulle2018simplified}. Specifically, we propose to generalize the well-known PnP-PGM~\cite{kamilov2017plug} and RED-SD~\cite{romano2017little} algorithms to their Bregman counterparts, PnP-BPGM and RED-BSD algorithms, by using the Bregman distance. We learn the corresponding artifact-removal operators by unfolding the iterations of our algorithms. Finally, we present the theoretical convergence analysis of PnP-BPGM and test our algorithms on Poisson linear inverse problems.  


\section{Background}
\subsection{Proximal Gradient Method}
 
 PGM can be interpreted as the Majorize-Minimization (MM) method for solving the composite optimization problem in~\eqref{invers_prob}. Each iteration of PGM can be expressed as a minimization of a quadratic majorizer
\begin{align}
\label{PGM_MM_iter}
\xbm^{k+1} =
\argmin_{\xbm \in \mathbb{R}^n} 
\left\{ \xbm^\Tsf \nabla f(\xbm^{k})
+ \frac{L}{2} \, \| \xbm - \xbm^{k} \|_2^2
+ g(\xbm) \right\},
\end{align}
\noindent 
where $f$ is assumed to have a $L$-Lipschitz continuous gradient. Eq.~\eqref{PGM_MM_iter} can also be expressed in the following form
\begin{subequations}
\begin{align}
\label{PGM_smo}
&\zbm^{k} = \xbm^{k} - \gamma \nabla f(\xbm^{k}) \\
&\xbm^{k+1} = \prox_{\gamma g}(\zbm^{k}),
\end{align}
\end{subequations}
\noindent where $0 < \gamma \leq 1/L$ is the step size and
\begin{align}
\label{Ec_prox}
\prox_{\gamma g}(\zbm) 
&\coloneqq \argmin_{\xbm \in \mathbb{R}^n} 
\left\{ \frac{1}{2} \left\| \zbm - \xbm \right\|_2^2 
+ \gamma g(\xbm) \right\}
\end{align}
\noindent 
is the proximal operator, which is well-defined for any proper, closed, and convex function $g$. 


\subsection{Using DL denoisers as priors}

The mathematical equivalence between proximal operator \eqref{Ec_prox} and Gaussian denoiser motivated denoiser-based iterative algorithms such as PnP \cite{venkatakrishnan2013plug}. In PnP, the proximal operator is replaced with an arbitrary Gaussian denoiser $\Dsf$.  Unlike PnP where the explicit regularizer $g$ is usually not known for a given denoiser, RED~\cite{romano2017little} seeks to form an explicit denoiser-based  regularizer $g(\xbm) = \tau \, \xbm^\Tsf (\xbm - \Dsf(\xbm))$, where $\tau > 0$ is the regularization parameter. When the denoiser is locally homogeneous and has a symmetric Jacobian \cite{reehorst2018regularization}, the gradient of the RED regularizer has a simple form $\nabla g(\xbm) =  \tau \, (\xbm - \Dsf(\xbm))$, which enables the usage of the traditional \emph{steepest descent (SD)} method for solving \eqref{invers_prob}. By leveraging the power of state-of-the-art DL-denoisers, such as DnCNN~\cite{Zhang.etal2017}, PnP/RED have achieved empirical success in many imaging applications~\cite{sreehari2016plug, brifman2016turning}. 

\subsection{The Bregman Distance}

Given a differentiable convex reference function $h$ defined on a closed convex set $C \subset \mathbb{R}^n$, the Bregman distance $B_h:  \mathrm{dom} \, h \times \mathrm{int \, dom}\, h \to [0, \infty)$ \cite{bregman1967relaxation} is defined by  
\begin{align}
B_h(\xbm; \ybm) 
&\coloneqq h(\xbm) - h(\ybm) 
- \nabla h(\ybm)^\Tsf (\xbm - \ybm).
\end{align}
The Bregman distance\footnote{Note that the Bregman distance is a pesudodistance, because it does not satisfy the triangle inequality, and is generally asymmetric.} is an extension of the classical squared Euclidean distance which is recovered when $h(\xbm) = 1/2 \| \xbm \|_2^2$. Other widely-used Bregman distance functions include the KL-divergence and the Itakura–Saito (IS) distance.

\section{Proposed Method} \label{sec:proposed_methods}

The path to Bregman-based proximal algorithm starts from observing that the quadratic majorization step in the classical PGM in eq.~\eqref{PGM_MM_iter} is equivalent to the following condition:
\begin{align}
\label{L_simple_condition}
\frac{L}{2} \, \| \xbm \|^2 - f(\xbm) \quad \text{is convex},
\end{align}
\noindent where the equivalence follows from the first-order convexity inequality. To bypass the Lipschitz gradient assumption, the work in \cite{bauschke2017descent, lu2018relatively} has proposed to generalize the condition in eq.~\eqref{L_simple_condition} by using a possibly non-quadratic reference function $h$
\begin{align}
\label{L_h_condition}
L \, h(\xbm) - f(\xbm) \quad \text{is convex}.
\end{align}
\noindent Such functions $f$ can be referred to as $L$-smooth relative to $h$. Then, by using the first-order convexity inequality, one can obtain a Bregman majorizer of the data-fidelity term
\begin{align}
\label{BPGM_MM_iter_1}
f(\xbm) \leq f(\xbm^{k}) + \nabla f(\xbm^{k})^\Tsf (\xbm - \xbm^{k}) + L \, B_h(\xbm, \xbm^{k}).
\end{align}
This inequality directly leads to the Bregman PGM (BPGM) method, which generalizes the classical PGM using a Bregman majorizer as
\begin{align}
\label{BPGM_MM_iter_2}
\xbm^{k+1} =
\argmin_{\xbm} 
\left\{
\xbm^\Tsf \nabla f(\xbm^{k})
+ L \, B_h(\xbm, \xbm^{k})
+ g(\xbm) \right\}.
\end{align}
The BPGM method shares the same structural splitting mechanism as the classical PGM \cite{teboulle2018simplified}, which allows one to express~\eqref{BPGM_MM_iter_2} as
\begin{subequations}
\label{BPGM_iter}
\begin{align}
\label{BPGM_MD}
&\zbm^{k}
= \nabla h^\ast \left( \nabla h - \gamma \nabla f \right) (\xbm^{k}) \\
&\xbm^{k+1} 
= \left( \nabla h + \gamma \, \partial g \right)^{-1} \nabla h ( \zbm^{k} )
\label{BPGM_iter_prox}
\end{align}
\end{subequations}
\noindent where $0 < \gamma \leq 1/L$ is the step size and $h^\ast$ denotes the Fenchel conjugate of $h$. Note that the PGM is a special case of the BPGM obtained by setting $h(\xbm) = 1/2 \| \xbm \|_2^2$. The first step of the BPGM in \eqref{BPGM_MD} is known as the Mirror Descent (MD) algorithm, which generalizes the classical GM. The second step is known as the \emph{left} Bregman proximal operator (BPO) defined as
\begin{align}
\label{Breg_prox}
\prox_{g}^{h}(\zbm) 
&\coloneqq \argmin_{\xbm} 
\left\{ B_h(\xbm, \zbm) + g(\xbm) \right\}.
\end{align}
Traditionally, the BPO is motivated from a computational perspective, e.g., Bregman projection onto the simplex with $h(\xbm)=\xbm^T \log(\xbm)$ is simpler than the corresponding classical proximal operator~\eqref{Ec_prox}. Moreover, selecting the reference function $h$ provides more flexibility depending on the problem settings \cite{bauschke2017descent, lu2018relatively, teboulle2018simplified}.

\begin{figure*}
	\begin{minipage}[t]{.49\textwidth}
		\begin{algorithm}[H]
			\caption{PnP-BPGM}\label{alg:bpgm}
			\begin{algorithmic}[1]
				\State \textbf{input: } $\xbm^0 \in \R^n$, $\ybm \in \R^m$, and $\gamma > 0$
				\For{$k = 1, 2, \dots$}
				\State $\xbm^{k+1} 
				= \Dsf_\theta \left(\nabla h^\ast \left( \nabla h - \gamma \nabla f \right) \right) (\xbm^{k})$
				\EndFor
			\end{algorithmic}
		\end{algorithm}%
		
	\end{minipage}
	\hspace{0.25em}
	\begin{minipage}[t]{.49\textwidth}
		\begin{algorithm}[H]
			\caption{RED-BSD}\label{alg:bred}
			\begin{algorithmic}[1]
				\State \textbf{input: } $\xbm^0 \in \R^n$, $\ybm \in \R^m$, and $\gamma > 0$
				\For{$k = 1, 2, \dots$}
				\State $\xbm^{k+1} = \nabla h^\ast \left( \nabla h - \gamma \left( \nabla f + \tau \, (I - \Dsf_\theta) \right) \right) (\xbm^{k})$
				\EndFor
			\end{algorithmic}
		\end{algorithm}%
	\end{minipage}
\end{figure*}

\begin{figure*}[h]
	\centering\includegraphics[width=0.99\textwidth]{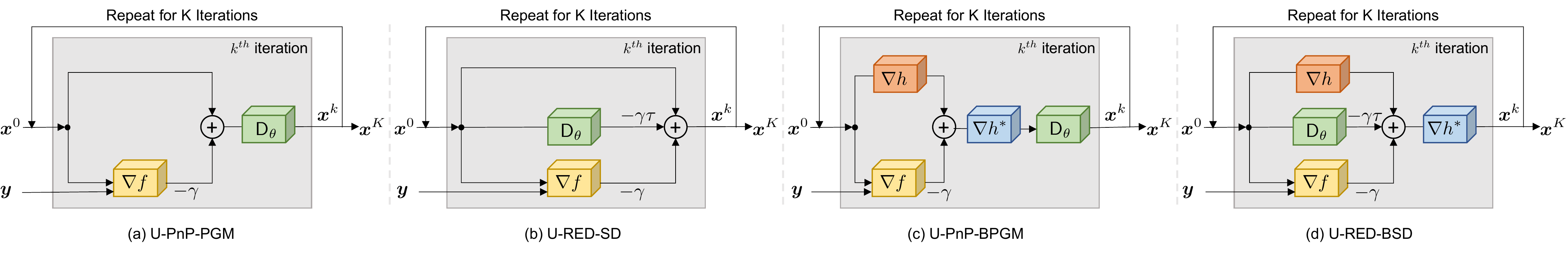}
	\caption{The proposed PnP-BPGM and RED-BSD methods replace the quadratic penalty in PnP-PGM and RED-SD by a more general Bregman distance. Both algorithms rely on data-driven regularizers obtained by training an artifact-removal operator $\Dsf_\theta$ via deep unfolding.}
	\label{Fig:unfold}
\end{figure*}

\subsection{Bregman PnP and RED Algorithms}

In this section, we propose two algorithms, PnP-BPGM and RED-BSD that extend existing two algorithms PnP-PGM and RED-SD, respectively. PnP-BPGM is obtained by replacing the BPO in \eqref{BPGM_iter_prox} with a DL network $\Dsf_\theta$ 
\begin{subequations}
\label{PnP_BPGM_iter}
\begin{align}
\label{PnP_BPGM_iter_for}
&\zbm^{k}
= \nabla h^\ast \left( \nabla h - \gamma \nabla f \right) (\xbm^{k}) \\
&\xbm^{k+1} 
= \Dsf_\theta (\zbm^{k}),
\end{align}
\end{subequations}
\noindent where $\theta$ are the learnable parameters that characterize the network $\Dsf_\theta$. Similarly, the Bregman variant of RED-SD is obtained as
\begin{align}
\label{RED_BPGM}
\xbm^{k+1} 
&= \nabla h^\ast \left( \nabla h 
- \gamma \left( \nabla f 
+ \tau \, (I - \Dsf_\theta) \right) \right) (\xbm^{k}),
\end{align}
\noindent where $I$ is a identity operator. When the assumptions for the existence of the explicit RED regularizer in~\cite{romano2017little} hold, then RED-BSD can be interpreted as the mirror descent algorithm. Note that the PnP-PGM \cite{venkatakrishnan2013plug} and RED-SD \cite{romano2017little} are recovered when $h(\xbm)= 1/2 \|\xbm\|^2$ and $\Dsf_\theta$ being a Gaussian denoiser. 

Algorithm~\ref{alg:bpgm} and Algorithm~\ref{alg:bred} summarize the proposed PnP-BPGM and RED-BSD algorithms. In this work, the regularizer $\Dsf_\theta$ is implemented using the deep unfolded strategy, so we refer to the proposed algorithms as unfolded PnP-BPGM (U-PnP-BPGM) and unfolded RED-BSD (U-RED-BSD). Similarly, the unfolded version of PnP-PGM, and RED-SD are referred as U-PnP-PGM and U-RED-SD. All four different unfolding architectures are shown in Fig.~\ref{Fig:unfold} and will be compared in the next section.

Recent work has explored the convergence properties of various PnP/RED algorithms~\cite{chen2018theoretical, sun2019online, ryu2019plug}. Similar results can be also established for both PnP-BPGM and RED-BSD. The following theorem presents the analysis of PnP-BPGM for a strongly convex function $f$ and a Lipschitz continuous operator $\Dsf_\theta$. While these assumptions are too strong for some applications, they provide the first steps for the broader analysis of Bregman PnP/RED methods.

\begin{theorem}
Assume $h$ be $\mu_h$-strongly convex with $L_h$-Lipschitz continuous gradient, and $f$ be $\mu_f$-strongly convex function with $L_f$-Lipschitz continuous gradient. Assume $\Dsf_\theta$ be an $M$-Lipschitz operator. Then, the iteration in \eqref{PnP_BPGM_iter} converges to a fixed point if
\begin{align}
M <  \frac{\mu_h \, (\mu_f + L_f)}{L_h \, L_f - \mu_h \, \mu_f}
\end{align}
and the step size 
$\frac{\mu_h}{\mu_f} \left( \frac{L_h}{\mu_h} - \frac{1}{M} \right)
< \gamma < 
\frac{\mu_h}{L_f} \left( 1 + \frac{1}{M} \right)$.
\end{theorem}

\section{Numerical illustration} \label{Sec:experiments}
\subsection{Poission linear inverse problem}
We empirically evaluated the proposed methods on Poisson linear inverse problems. Poisson noise is a signal dependent noise whose negative log-likelihood function results in the following data-fidelity term and its gradient
\begin{subequations}
\begin{align}
\label{poisson_function}
f(\xbm) &= \mathbf{1}^\Tsf \Abm \xbm - \ybm^\Tsf \log\left(\Abm \xbm \right) + \mathbf{1}^\Tsf \log(\ybm !) \\
\nabla f(\xbm) &= \Abm^\Tsf \left( \mathbf{1} - \ybm \oslash (\Abm \xbm) \right)
\end{align}
\end{subequations}
\noindent where $\mathbf{1}$ is a vector of ones, and $\oslash$ denotes element-wise division.

Classical algorithms for solving Poisson linear inverse problems include the Richardson–Lucy (RL) algorithm and transform-based methods \cite{harmany2011spiral, sarder2006deconvolution, starck2007astronomical, dey2006richardson, makitalo2010optimal, dupe2009proximal}. Several ADMM-based algorithms were proposed that handle the data fidelity via its proximal operator \cite{figueiredo2010restoration, rond2016poisson}. 
In \cite{bauschke2017descent} it was showed that by using the Burg’s entropy as a reference function $h(\xbm) = - \mathbf{1}^\Tsf \log(\xbm)$, one can satisfy \eqref{L_h_condition} with $L \geq \| \ybm \|_1$. Therefore, using \eqref{PnP_BPGM_iter} we can obtain the following simple iteration for PnP-BPGM
\begin{subequations}
\begin{align}
&\zbm^{k} = \frac{\xbm^{k}}{\mathbf{1} + \gamma \, \xbm^{k} \odot \nabla f(\xbm^{k}) } \\
&\xbm^{k+1} = \Dsf_\theta( \zbm^{k} ),
\end{align}
\end{subequations}
\noindent where $\odot$ is the element-wise multiplication. It can be shown that the backward operator is related to inverse Gamma scale estimator. Similarly, RED-BSD in \eqref{RED_BPGM} can be simplified to
\begin{align}
\xbm^{k+1} &= \frac{\xbm^{k}}{\mathbf{1} 
+ \gamma \, \xbm^{k} \odot ( \nabla f + \tau \, (I - \Dsf_\theta ) )(\xbm^{k})}.
\end{align}

\begin{table*}[t!]
	\centering
	\caption{The PSNR (dB) results of different methods on the testing images with different peaks and kernels.}
	\begin{tabular}{ |p{2.3cm}|p{0.74cm}|p{0.74cm}|p{0.74cm}| p{0.74cm}|p{0.74cm}|p{0.74cm}|p{0.74cm}|p{0.74cm}|p{0.74cm}|p{0.74cm}|p{0.74cm}|p{0.74cm}|p{1.1cm}|}
		\hline
		Method &1 &2 &3 &4 &5 &6 &7 &8 &9 &10 &11 &12 & Average\\ 
		\hline \hline
		\multicolumn{14}{|c|}{Uniform kernel, peak = 8}\\
		\hline
		Corrupted & 11.70 & 11.13 & 11.74 & 11.59 & 11.61 & 9.56 & 11.81 & 11.80 & 11.82 & 11.63 & 12.04 & 11.91 & 11.53 \\
		U-Net & 20.89 & 23.11 & \textbf{21.28} & 20.79 & 19.79 & 19.15 & 19.63 & 24.76 & 21.96 & \textbf{22.70} & 23.37 & 22.65 & 21.67 \\
		U-PnP-PGM & 19.57 & 22.74 & 20.89 & 20.59 & 19.20 & 19.15 & 19.04 & 24.38 & 21.79 & 22.43 & 23.19 & 22.44 & 21.28 \\
		U-RED-SD & 20.38 & 22.74 & 20.74 & 20.29 & 18.81 & 19.00 & 18.90 & 24.44 & 21.81 & 22.38 & 23.32 & 22.41 & 21.27 \\
		U-PnP-BPGM & \textbf{21.00} & \textbf{24.12} & 21.27 & 20.72 & 20.10 & 19.17 & \textit{\textbf{19.81}} & \textbf{25.21} & \textbf{22.13} & 22.65 & \textbf{23.82} & 22.62 & \textbf{21.89} \\
		U-RED-BSD & 20.97 & 23.97 & 21.14 & \textbf{20.82} & \textbf{20.25} & \textbf{19.28} & 19.78 & 25.08 & \textbf{22.13} & \textbf{22.70} & 23.75 & \textbf{22.66} & 21.88 \\ 
		\hline
		\multicolumn{14}{|c|}{Uniform kernel, peak = 32}\\
		\hline
		Corrupted & 16.14 & 16.08 & 16.51 & 16.25 & 15.76 & 13.97 & 15.81 & 17.12 & 16.68 & 16.77 & 17.13 & 16.95 & 16.26 \\
		U-Net & 21.50 & 24.66 & 22.12 & 21.71 & 21.01 & 19.97 & 20.23 & 26.19 & 22.56 & 23.38 & 24.67 & 23.40 & 22.62 \\
		U-PnP-PGM & 20.90 & 23.76 & 22.03 & 21.54 & 20.54 & 19.71 & 19.85 & 25.37 & 22.22 & 23.26 & 24.01 & 23.32 & 22.21 \\
		U-RED-SD & 21.37 & 24.13 & 21.92 & 21.41 & 20.75 & 19.88 & 20.17 & 25.67 & 22.40 & 23.35 & 24.25 & 23.40 & 22.39 \\
		U-PnP-BPGM & \textbf{21.58} & 25.01 & 22.15 & \textbf{21.81} & \textbf{21.64} & \textbf{20.23} & \textbf{20.57} & 26.33 & 22.64 & \textbf{23.45} & 24.90 & \textbf{23.46} & \textbf{22.81} \\
		U-RED-BSD & 21.57 & \textbf{25.04} & \textbf{22.17} & 21.64 & 21.48 & 20.22 & 20.34 & \textbf{26.44} & \textbf{22.65} & 23.35 & \textbf{24.91} & 23.41 & 22.77 \\
		\hline
		\multicolumn{14}{|c|}{Gaussian kernel, peak = 8}\\
		\hline
		Corrupted & 11.98 & 11.25 & 12.01 & 11.86 & 12.01 & 9.71 & 12.32 & 11.89 & 11.89 & 11.79 & 12.17 & 12.07 & 11.75 \\
		U-Net & 21.72 & 24.92 & 22.06 & 21.55 & 22.17 & 20.87 & 21.27 & 25.60 & 22.22 & 22.97 & 24.63 & 23.08 & 22.76 \\
		U-PnP-PGM & 21.01 & 23.97 & 21.70 & 21.41 & 20.74 & 19.72 & 20.32 & 25.18 & 22.17 & 22.72 & 24.17 & 22.87 & 22.16 \\
		U-RED-SD & 21.18 & 23.30 & 21.88 & 21.26 & 20.65 & 19.79 & 20.15 & 24.86 & 21.96 & 22.97 & 23.69 & 23.03 & 22.06 \\
		U-PnP-BPGM & \textbf{22.30} & 24.60 & \textbf{22.48} & \textbf{21.78} & \textbf{22.44} & 19.23 & \textbf{21.92} & \textbf{26.03} & \textbf{22.48} & \textbf{23.90} & \textbf{24.49} & \textbf{23.60} & \textbf{22.94} \\
		U-RED-BSD & 22.22 & \textbf{24.62} & 22.17 & 21.72 & 22.27 & \textbf{19.61} & 21.61 & 25.76 & 22.37 & 23.79 & 24.37 & \textbf{23.60} & 22.84 \\
		\hline
		\multicolumn{14}{|c|}{Gaussian kernel, peak = 32}\\
		\hline
		Corrupted & 17.06 & 16.62 & 17.30 & 17.17 & 17.05 & 14.45 & 17.19 & 17.51 & 17.04 & 17.35 & 17.57 & 17.55 & 16.99 \\
		U-Net & 22.63 & 26.74 & 23.13 & 23.13 & 23.83 & \textbf{21.69} & 22.51 & 27.14 & 22.89 & 24.00 & 25.95 & 24.12 & 23.98 \\
		U-PnP-PGM & 22.12 & 24.50 & 23.61 & 23.10 & 22.54 & 19.53 & 21.81 & 26.03 & 22.60 & 24.27 & 24.77 & 24.22 & 23.26 \\
		U-RED-SD & 22.15 & 25.43 & 23.07 & 23.14 & 22.86 & 21.29 & 21.92 & 26.55 & 22.80 & 24.27 & 25.31 & 24.24 & 23.58 \\
		U-PnP-BPGM & \textbf{23.41} & \textbf{26.85} & \textbf{23.79} & \textbf{23.54} & 24.41 & 20.82 & \textbf{23.30} & 27.86 & \textbf{23.13} & \textbf{25.03} & 26.03 & \textbf{24.83} & \textbf{24.42} \\
		U-RED-BSD & 23.12 & 26.79 & 23.41 & 23.27 & \textbf{24.46} & 21.02 & 23.05 & \textbf{27.88} & 23.11 & 24.96 & \textbf{26.04} & 24.75 & 24.32 \\
		\hline
	\end{tabular}
	\label{tab:my_label}
\end{table*}
\begin{figure*}[h]
	\centering\includegraphics[width=\textwidth]{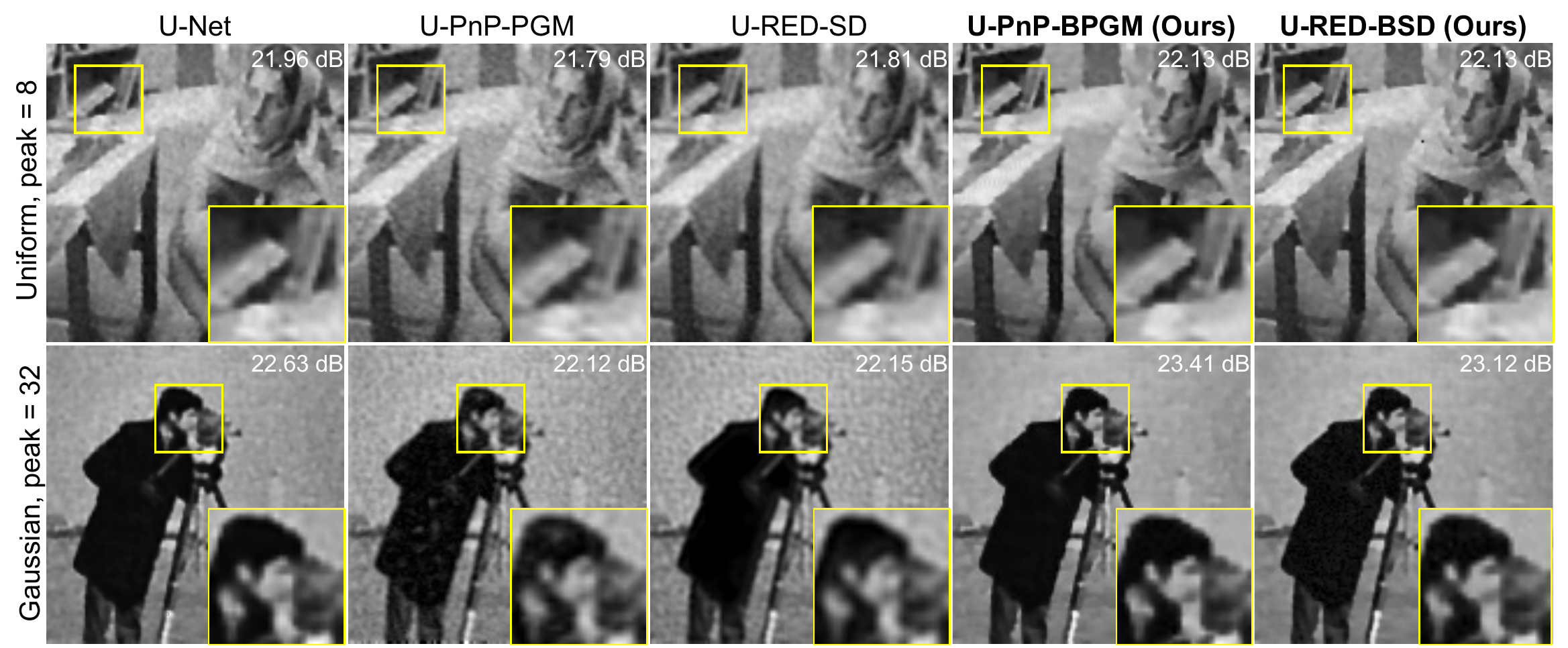}
	\caption{Examples of image reconstruction results on \textit{Babara} (top) and \textit{Cameraman} (bottom) obtained by U-Net, U-PnP-PGM, U-RED-SD, U-PnP-BPGM, and U-RED-BSD. The first row is corresponding to the noise peak 8 with uniform kernel, and the second row is noiser peak 32 with Gaussian kernel. Each reconstruction is labeled with its PSNR (dB) value with respect to the Ground-truth image. Visual differences are highlighted using the rectangles drawn inside the images. Note U-PnP-BPGM and U-RED-BSD shows close performance one to another, outperforming other methods and providing the best visual results by recovering sharp edges and removing artifacts.}
	\label{Fig:vis}
\end{figure*}

\subsection{Image Deblurring with Possion noise}

We demonstrate the ability of our proposed algorithms PnP-BPGM and RED-BSD over their traditionally counterparts PGM and RED on Poisson linear inverse problems.  We focus on image deblurring, where the forward model $\Abm$ corresponds to the blurring operator. Specifically, we follow a similar settings in~\cite{figueiredo2010restoration, rond2016poisson}, and test our algorithms for Poisson noise with peaks 8 and 32 using two different blur kernels of size 9 by 9: (1) a Gaussian kernel with $\sigma = 1.6$, and (2) a  uniform kernel, respectively. All the methods compared are trained in an unfolding fashion as illustrated in Fig.~\ref{Fig:unfold}, where the end-to-end training seeks to compute the trainable parameters in $\Dsf_\theta$ by minimizing the $\ell_2$ loss function between network output $\{\xbm^K\}$ and the ground-truth $\{\xbm\}$ over all training samples. We set $\xbm^0$ using the raw measurements $\ybm$ with a small white Gaussian perturbation. We unfold each algorithm with $ K = 100 $ iterations for stable performance,  where in each iteration, the network $\Dsf_\theta$ is realized using a $7-$layer DnCNN~\cite{Zhang.etal2017} with shared weights across all iterations.  The step-size parameter $\gamma$ and the regularization parameter $\tau$ in RED and BRED are set as a learnable parameters, initialized with $\gamma = 5\times 10^{-1}$ and $\tau = 1 \times 10^{-3}$. As a reference, we also report the image reconstruction performance of the end-to-end learning method where U-Net~\cite{Ronneberger.etal2015} is trained end-to-end in the usual supervised fashion using the $\ell_2$-loss~\cite{Jin.etal2017a}. All networks are trained on public dataset BSD400 for 400 epochs, using the Adam solver~\cite{Kingma.Ba2015} with an initial learning rate $1\times 10^{-4}$. We select the models that achieved the best performance on the validation dataset BSD68. At test time, Set12 dataset is used to evaluate the performance of each algorithm.

The numerical results on the test dataset Set12  with respect to two  scenarios are summarized in Table~\ref{tab:my_label}. Test images used for the quantitative performance labeled from 1 to 12 are: \textit{Cameraman}, \textit{House},  \textit{Pepper},  \textit{Starfish},  \textit{Butterfly},  \textit{Plane},  \textit{Parrot}, \textit{Lena}, \textit{Barbara}, \textit{Boat}, \textit{Artist}, \textit{Room}. For each image, the highest PSNR in each scenario is highlighted. We observe that the performances of U-PnP-BPGM and U-RED-BSD are very close to one another, providing the best performance compared to all the other methods, outperforming U-PnP-PGM and U-RED-SD. Fig.~\ref{Fig:vis} shows visual examples for two images from Set12 in two different settings, uniform kernel with peak 8 (top) and Gaussian kernel with peak 32 (bottom). Note that both U-PnP-PGM and U-RED-SD yield similar visual recovery performance with artifacts remaining in the images, U-PnP-BPGM and U-RED-BSD show much better reconstruction performance in removing artifacts and noise. The enlarged regions in the image suggest that U-PnP-BPGM and U-RED-BSD better recover the fine details and sharper edges compared to their counterparts and U-Net.

\section{Conclusion} \label{sec:conclusion}

This paper proposes generalizing plug-and-play priors (PnP) and regularization by denoising (RED) beyond squared Euclidean distance using the Bregman distance. The proposed Bregman-based methods are motivated by the recent progress in optimization, that have the potential to better align to specific non-Euclidean geometry of the loss function. Our numerical results show the potential of the proposed methods in Poisson linear inverse problems. This work can be considered as a first step towards extending widely-used PnP/RED to problems where there is a benefit of using non-Euclidean formulations of proximal and projection operators.





\end{document}